\documentclass[journal]{IEEEtran}
\makeatletter\if@twocolumn\PassOptionsToPackage{switch}{lineno}\else\fi\makeatother

\usepackage{graphicx}
\usepackage[T1]{fontenc}
\usepackage{wrapfig}

\usepackage{scalerel}
\usepackage{tikz}
\usetikzlibrary{svg.path}
\usepackage[switch]{lineno}
\usepackage{float}

\definecolor{orcidlogocol}{HTML}{A6CE39}
\tikzset{
  orcidlogo/.pic={
    \fill[orcidlogocol] svg{M256,128c0,70.7-57.3,128-128,128C57.3,256,0,198.7,0,128C0,57.3,57.3,0,128,0C198.7,0,256,57.3,256,128z};
    \fill[white] svg{M86.3,186.2H70.9V79.1h15.4v48.4V186.2z}
                 svg{M108.9,79.1h41.6c39.6,0,57,28.3,57,53.6c0,27.5-21.5,53.6-56.8,53.6h-41.8V79.1z M124.3,172.4h24.5c34.9,0,42.9-26.5,42.9-39.7c0-21.5-13.7-39.7-43.7-39.7h-23.7V172.4z}
                 svg{M88.7,56.8c0,5.5-4.5,10.1-10.1,10.1c-5.6,0-10.1-4.6-10.1-10.1c0-5.6,4.5-10.1,10.1-10.1C84.2,46.7,88.7,51.3,88.7,56.8z};
  }
}
\newcommand\orcidicon[1]{\href{https://orcid.org/#1}{\mbox{\scalerel*{
\begin{tikzpicture}[yscale=-1,transform shape]
\pic{orcidlogo};
\end{tikzpicture}
}{|}}}}

\newif\ifpreprint
\preprinttrue

\ifpreprint
\definecolor{revised}{rgb}{0,0,0} 
\else
\definecolor{revised}{rgb}{0.1,0.1,0.6} 
\fi

\definecolor{huai}{rgb}{0.1,0.1,0.6}
\definecolor{hai}{rgb}{0.1,0.1,0.6}
\definecolor{hung}{rgb}{0.06,0.2,0.65}
\definecolor{ty}{rgb}{0.1,0.1,0.6}

\newcommand*\circled[1]{\tikz[baseline=(char.base)]{
            \node[shape=circle,draw,inner sep=1pt] (char) {#1};}}
            
\ifCLASSINFOpdf
\else
\fi
\usepackage{amsmath}
\usepackage{algorithm}
\usepackage{algpseudocode}

\include{textbox} 

\usepackage{url,multirow,morefloats,floatflt,cancel,tfrupee}

\ifpreprint
\usepackage[letterpaper, left=0.6in, right=0.5in, bottom=0.6in, top=0.6in]{geometry}
\else
\usepackage[letterpaper, left=0.7in, right=0.7in, bottom=0.6in, top=0.6in]{geometry}
\fi

\usepackage{adjustbox}
\usepackage{hyperref} 
\usepackage{kbordermatrix} 

\makeatletter

\AtBeginDocument{\@ifpackageloaded{textcomp}{}{\usepackage{textcomp}}}
\makeatother
\usepackage{colortbl}
\usepackage{xcolor}
\makeatletter

\def\mcWidth#1{\csname TY@F#1\endcsname+\tabcolsep}

\def\cAlignHack{\rightskip\@flushglue\leftskip\@flushglue\parindent\z@\parfillskip\z@skip}
\def\rAlignHack{\rightskip\z@skip\leftskip\@flushglue \parindent\z@\parfillskip\z@skip}

\@ifundefined{etal}{}{}

\usepackage{ifxetex}
\ifxetex\else\if@twocolumn\@ifpackageloaded{stfloats}{}{\usepackage{dblfloatfix}}\fi\fi

\AtBeginDocument{
\expandafter\ifx\csname eqalign\endcsname\relax
\def\eqalign#1{\null\vcenter{\def\\{\cr}\openup\jot\m@th
  \ialign{\strut$\displaystyle{##}$\hfil&$\displaystyle{{}##}$\hfil
      \crcr#1\crcr}}\,}
\fi
}

\AtBeginDocument{%
  \@ifpackageloaded{endfloat}%
   {\renewcommand\efloat@iwrite[1]{\immediate\expandafter\protected@write\csname efloat@post#1\endcsname{}}}{\newif\ifefloat@tables}%
}%

\def\BreakURLText#1{\@tfor\brk@tempa:=#1\do{\brk@tempa\hskip0pt}}
\let\lt=<
\let\gt=>
\def\processVert{\ifmmode|\else\textbar\fi}

\@ifundefined{subparagraph}{
\def\subparagraph{\@startsection{paragraph}{5}{2\parindent}{0ex plus 0.1ex minus 0.1ex}%
{0ex}{\normalfont\small\itshape}}%
}{}

\newcommand\role[1]{\unskip}
\newcommand\aucollab[1]{\unskip}
  
\@ifundefined{tsGraphicsScaleX}{\gdef\tsGraphicsScaleX{1}}{}
\@ifundefined{tsGraphicsScaleY}{\gdef\tsGraphicsScaleY{.9}}{}
\def\checkGraphicsWidth{\ifdim\Gin@nat@width>\linewidth
	\tsGraphicsScaleX\linewidth\else\Gin@nat@width\fi}

\def\checkGraphicsHeight{\ifdim\Gin@nat@height>.9\textheight
	\tsGraphicsScaleY\textheight\else\Gin@nat@height\fi}

\def\fixFloatSize#1{}
\let\ts@includegraphics\includegraphics

\def\inlinegraphic[#1]#2{{\edef\@tempa{#1}\edef\baseline@shift{\ifx\@tempa\@empty0\else#1\fi}\edef\tempZ{\the\numexpr(\numexpr(\baseline@shift*\f@size/100))}\protect\raisebox{\tempZ pt}{\ts@includegraphics{#2}}}}

\AtBeginDocument{\def\includegraphics{\@ifnextchar[{\ts@includegraphics}{\ts@includegraphics[width=\checkGraphicsWidth,height=\checkGraphicsHeight,keepaspectratio]}}}

\DeclareMathAlphabet{\mathpzc}{OT1}{pzc}{m}{it}

\def\URL#1#2{\@ifundefined{href}{#2}{\href{#1}{#2}}}

\def\UrlOrds{\do\*\do\-\do\~\do\'\do\"\do\-}%
\g@addto@macro{\UrlBreaks}{\UrlOrds}

\edef\fntEncoding{\f@encoding}

\makeatother

\newif\ifmultipleabstract\multipleabstractfalse%
\makeatletter
\AtBeginDocument{\@ifpackageloaded{longtable}{%
\def\LT@makecaption#1#2#3{%
  \LT@mcol\LT@cols c{\hbox to\z@{\hss\parbox[t]\LTcapwidth{%
    \sbox\@tempboxa{#1{#2: } #3}%
    \ifdim\wd\@tempboxa>\hsize
      #1{#2: }\textsc{#3}%
    \else
      \hbox to\hsize{\hfil\box\@tempboxa\hfil}%
    \fi
    \endgraf\vskip\baselineskip}%
  \hss}}}
}{}}
\makeatother

\begin{document}

%
\title{A Sequential Metamorphic Testing Framework for Understanding Automated Driving Systems}

\ifpreprint

\author{ {Quang-Hung~Luu}, {Huai~Liu}, {Tsong~Yueh Chen} and {Hai~L.~Vu}
\thanks{Corresponding author(s): Huai~Liu.\\
Quang-Hung~Luu, Huai~Liu, Tsong~Yueh Chen are with the Department of {Computing Technologies}, Swinburne University of Technology,
Hawthorn, VIC 3122, Australia (e-mail: hluu@swin.edu.au;
hliu@swin.edu.au; tychen@swin.edu.au).
Hai L. Vu, Quang-Hung Luu are with the Department of Civil Engineering, Monash University,
Clayton, VIC 3800, Australia (e-mail: hai.vu@monash.edu; hung.luu@monash.edu).}
}

\fi

%
%


\ifpreprint
\markboth{Preprint}{}
\fi 
        
%





\maketitle 
%
\IEEEpeerreviewmaketitle

\begin{abstract}
Automated driving systems (ADS) are expected to be reliable and robust against a wide range of driving scenarios. Their decisions, first and foremost, must be well understood. Understanding a decision made by ADS is a great  challenge, because it is not straightforward to tell whether the decision is correct or not, and how to verify it systematically. In this paper, a Sequential MetAmoRphic Testing (\textsc{Smart}) framework is proposed based on metamorphic testing, a mainstream software testing approach. In metamorphic testing, metamorphic groups are constructed by selecting multiple inputs according to the so-called metamorphic relations, which are basically the system's necessary properties; the violation of certain relations by some corresponding metamorphic groups implies the detection of erroneous system behaviors. The proposed framework makes use of sequences of metamorphic groups to understand ADS behaviors, and is applicable without the need of ground-truth datasets. To demonstrate its effectiveness, the framework is applied to test three ADS models that steer an autonomous car in different scenarios with another car either leading in front or approaching in the opposite direction. The conducted experiments reveal a large number of undesirable behaviors in these top-ranked deep learning models in the scenarios. These counter-intuitive behaviors are associated with how the core models of ADS respond to different positions, directions and properties of the other car in its proximity. Further analysis of the results helps identify critical factors affecting ADS decisions and thus demonstrates that the framework can be used to provide a comprehensive understanding of ADS before their deployment.
\end{abstract}

\section{Introduction}\label{section:intro}

The advancement of automated driving systems (ADS) in the last few years has been inevitably revolutionizing the transportation sector. At the full automation level, the ADS will take over all driving tasks without the attention of human, offering a great deal of convenience for drivers.
The autonomous vehicles (AVs) underpinned by ADS promise to decrease fuel consumptions, reduce congestions, provide a safer transportation and increase the mobility \cite{anderson2014}. 
In addition to the modular ADS, the end-to-end approach has emerged as an alternative trend in {\color{revised}ADS research \cite{yurtsever2020,tampuu2020}.}
Having said that, the current systems are not as reliable and robust as expected. Numerous fatal crashes due to the failure of ADS have been reported. In 2016, the first driver was killed while operating his Tesla Model S in Autopilot (Tesla's ADS) mode just three months after the first crash of Google's self-driving cars into a bus in Mountain View in February \cite{nyholm2018a}. Within two years from these accidents, the first pedestrian was killed in Tempe, Arizona by an Uber test vehicle \cite{levin2018}. More recently, in May 2021, a crash caused by the Tesla Model 3 killed its driver while the Autopilot mode was in operating in Fontana, California \cite{cbsla2021}. These fatal accidents raise scepticisms for the safety of the state-of-the-art ADS and prohibit AVs from being accepted to be used on-road. To gain the trustworthiness, decisions made by ADS must be well understood and explainable before they can be deployed \cite{zablocki2021, huang2020, benchekroun2020,rajabli2021}. 

Obtaining a comprehensive understanding of the ADS is a great challenge, because it is often not straightforward to determine whether a decision made by the ADS is correct or not. For example, for the same situation of a simple car-following scenario involving an autonomous vehicle and another car appearing at a certain distance ahead, system A may drive the AV to steer rightward by $10^o$, whereas system B may guide it to go straight.
Under many circumstances, it is unable to determine which decision is more correct, as well as whether both decisions are correct or incorrect. From the testing perspective, to understand and then explain the decisions made by a system, several tests are required alongside a mechanism for determining whether or not the tests have passed (for making a correct decision) or failed (for making a wrong decision). Such a mechanism, which is referred to as the test oracle in software testing, often does not exist in the state-of-the-art ADS due to two main reasons. Firstly,  people may have different views on which decision can be considered correct {\color{revised}\cite{awad2018}}. For example, it is debatable about whether the ADS, while foreseeing a potential crash, should give priority to the individual safety of the driver or to the ``overall'' safety of other road users {\color{revised}\cite{nyholm2018a,awad2018}}. Such a choice significantly affects the algorithm implemented and thus makes the judgement hard to be conclusive. Secondly, most systems adopt the deep neural networks (DNN) for object and event detection and responses (OEDR), among other tasks. While DNN is advantageous in its strong learning capability without supervision which allows them to process a great amount of data \cite{goodfellow2016}, the decision made by such systems is unexplainable due to several reasons, such as the bias in the datasets used for training, the distinct patterns captured by a DNN architecture \cite{zablocki2021, huang2020, benchekroun2020, rajabli2021} and even the bugs existing in the system \cite{garcia2020}. 

These testing challenges can be alleviated using the Metamorphic Testing (MT) technique. 
MT has been successfully applied to test a wide range of sophisticated systems \cite{chen2017, xie2011,dwarakanath2018, zhang2019,luu2021}. 
Its applications have been expanded beyond testing to other purposes such as system understanding, fault localization, program repair, and cybersecurity \cite{segura2016, chen2017,chan2021tnnls,chen2021acm}. 
The emergence of AVs has triggered the application of  MT in assuring their safety with a number of state-of-the-art frameworks including DeepTest \cite{tian2018} and DeepRoad \cite{zhang2018deeproad}.
The use of MT helped identify thousands of erroneous behaviors in different deep learning models driving the ADS.
However, deeper and more profound information is yet to be explored. More importantly, there does not exist a framework facilitating a comprehensive system understanding of ADS. 
To fill this gap, a novel and effective approach is proposed for uncovering deeper information in ADS, which has been overlooked by existing MT-based testing techniques. In this study, a new framework is proposed to provide a systematic understanding of {\color{revised}ADS.}
The key contributions of the present paper are as follows. 
\begin{itemize}
	\item Proposal of a novel framework (\textsc{Smart}) to make use of the relationships between metamorphic groups (that is, groups of multiple related inputs) to uncover deeper information for the understanding of ADS' decisions.
	\item Application of the framework to test three well-known ADS models predicting steering angle, which helps reveal a larger number of undesirable behaviors of these models.
	\item Development of an open source software to support the implementation of the \textsc{Smart} framework.
\end{itemize}

The rest of the paper is organized as follows. In Section~\ref{section:review}, the background of MT, the state of the art for MT in testing the ADS and the gap in understanding the ADS decisions are presented. Then the motivation and description of framework are presented in Section~\ref{section:framework}. 
The experiment setting is detailed in Section~\ref{section:experiment}, which includes deep learning models to be tested, the metrics for checking undesirable behaviors of steering angles and the generation of data. Experimental results are reported in Section~\ref{section:results} before their insights are discussed in Section~\ref{section:insights}.
Finally, the paper is concluded in Section~\ref{section:conclusion}.

\section{Background and Literature Review}\label{section:review}

In this section, the concept of MT is briefly introduced before reviewing key related studies in testing ADS. The review then points to the importance of a systematic understanding of ADS, the research direction that has not been fully explored, and the reason why MT is an ideal technique to address it.


\subsection{Preliminary of metamorphic testing}

MT differs from traditional testing techniques in the sense that it makes use of the relations between multiple executions of the systems under test. These relations, referred to as metamorphic relations (MRs), are derived from necessary properties of the system. Once an MR is violated, it is known that the system is faulty. For example, assume that $G$ is a graph having 100 nodes and a program $P$ finding shortest paths $P(G,a,c)$  between any two nodes $a$ and $c$ in $G$. A naive process to verify this program needs to cover against $100!$ possible paths. However, based on the domain knowledge of the shortest path in a graph, one can derive the MR ``$|P(G,a,c)|\le |P(G,a,b)|+|P(G,b,c)|$'' with an arbitrary (connected) node $b$ where $|\cdot|$ denotes the length of shortest path. If the program $P$ does not uphold this MR for a set of nodes $(a,b,c)$ of the graph $G$, it is known to be faulty.

MT can provide an alternative mechanism to the test oracle for verifying and validating a system {\color{revised}\cite{segura2018b}}. It can also help generate new test cases automatically and effectively.
The basic steps for implementing MT are as follows:
\begin{enumerate}
\item Necessary properties of the target system are identified in the form of relations among multiple inputs and corresponding expected outputs, namely metamorphic relations (MRs).
\item From a set of given inputs (referred to as {\it source inputs}), a new set of inputs (namely, {\it follow-up inputs}) is constructed using the MR. Both sets of inputs form a metamorphic group (MG). Different MGs can be constructed using the same MR. 
\item The {\it source outputs} and the {\it follow-up outputs} are obtained after executing  the program under test with an MG. In general, the source output may also be involved in the construction of the follow-up input.
\item The outputs are checked against the MR to determine whether it is violated or not. The violation of an MR implies that the system is faulty.
\end{enumerate}

As described above, the violation of MR (thus, the failure of the system) is determined by evaluating decisions for individual MGs of input. In the previous graph example, from the source input $(G,a,c)$, two follow-up inputs are generated, that is, $(G,a,b)$ and $(G,b,c)$, to form an MG. Multiple executions using $P$ for this MG give us one source output $|P(G,a,c)|$ and two follow-up outputs $|P(G,a,b)|$ and $|P(G,b,c)|$, respectively. These outputs are then validated against the MR to check the correctness of program or system under test. 

Let us look at another example of autonomous vehicles. Assume that a system controlling steering angle (SA) based on images captured by a car's camera is tested. An MR is defined as: the system's decisions are robust against different weathers, that is, the steering angle shall not change much, with respect to changes of weathers \cite{tian2018}. An image capturing a driving scenario in the sunny weather is adopted as a source input. The follow-up input can be generated by transforming it to a new image, keeping the same scenario but under a snowy weather. If the SA for the snowy weather significantly differs the SA for the normal day, the relation does not hold; hence, system is faulty in the sense that it is not robust enough against weathers.


\subsection{Testing the ADS with metamorphic testing}

In the existing literature, two testing strategies have been commonly applied to test the ADS: the model-based and the MT approaches. The former approach adopts formal model specifications or cross-model references to determine system faults. They are defined by means of ``safety cages'' to determine the operational bounds \cite{kuutti2019},  ``cross referencing oracles'' defined by the output of majority of various models in combination with the notion of neuron coverage to judge incorrect corner-case behaviors \cite{pei2017} or frontier boundaries in which the system starts responding undesirably \cite{riccio2020}. This approach suffers from limitations of efficiency, effectiveness and flexibility in applications to complex ADS systems \cite{rajabli2021}. 

Recently, the MT approach has gained noticeable successes in testing complex ADS systems \cite{tian2018, zhang2018deeproad, xie2018a, zhou2019}. Tian {\it et~al.} \cite{tian2018} conducted a pioneer study where an MT-based framework, namely DeepTest, was developed. DeepTest used the image dataset from Udacity challenge as the source inputs. They then generated follow-up inputs from the dataset to mimic different weather and distortable conditions. Each pair of the source (original) image and the follow-up (transformed) image forms an MG. Their proposed MR requires the difference between predicted SA for each MG to be smaller than a certain predefined threshold; otherwise, the system is considered as faulty. They successfully revealed erroneous behaviors of three ADS models in handling transformed inputs, including rains, fogs, lightings or signal degradations.

Following the success of DeepTest framework,  Zhang {\it et~al.} \cite{zhang2018deeproad} devised DeepRoad, an offline framework that employed generative adversarial networks (GAN)-based techniques to generate synthesized driving scenes, and obtain inputs that are diverse and more realistic. Similar to Tian {\it et~al.} \cite{tian2018}, they constructed individual MGs, each composed of a source driving image and a follow-up image with the scene transformed by a GAN tool. They used MRs being defined such that differences between predicted SAs for individual MGs are negligible. DeepRoad helped uncover thousands of inconsistent behaviors in three ADS models, and at the same time, is potential to be used for validating the images transformed by GAN. 

Recently, MT has been combined with other techniques to tackle more challenging testing tasks for the ADS. Zhou and Sun \cite{zhou2019}  adopted fuzzing-based MT techniques to examine the reliability of LiDAR obstacle-perception module of Apollo, a well-known multi-platform for self-driving cars. They found several issues with Apollo. For example, from the source inputs containing 3D cloud points, they generated follow-up inputs by adding 1,000 random data points outside the region of interest (ROI) of the Apollo. It was found after comparing the executions for these MGs that surrounding cars currently in the ROI would no longer be detectable. In another case, having a follow-up input imitating a small insect flying 100 meters away from the AV, which is a simulated vehicle underpinned by the Apollo platform, may confuse the system in detecting the pedestrian moving in front of that car. DeepHunter is another framework integrating MT with fuzzing \cite{xie2018a} and is applicable to ADS. In DeepHunter, MRs were combined with multiple extensible coverage criteria as a strategy to generate new test cases that preserve the test semantics.


\subsection{Understanding of ADS and its research gaps}

Understanding helps users gain the in-depth knowledge about the system \cite{zhou2018}. It is the deeper layer of information that is explainable by human but is not reflected in the test score counting the numbers of passed and failed test cases \cite{zablocki2021}. 
On the other hand, testing is a major approach to verification checking whether the software meets the predefined specifications. Understanding and testing are strongly related in a feedback-loop fashion where understanding helps improve the adequacy of testing and testing facilitates understanding.
However, there is a major difference between them.
A good ``test score'' in the testing to reveal faults is not enough to guarantee a safe system and increase the usability. That is to say, ``incorrect'' programs may still be useful whilst non-faulty program may be impractical \cite{lehman1980,zhou2018}. 

A number of studies  \cite{benchekroun2020,zablocki2021} have pointed out that parties involved in the development and use of ADS all have benefits in understanding its behaviors. Understanding is the first step to help end-users build up their trust. For legal authorities, the understanding helps resolve problems with the liability, especially in regulating the safety requirements. By having a sound understanding of the ADS, designers and manufacturers are able to know their ADS' limitations and way to improve it. Furthermore, it helps engineers to build better versions based on revealed corner cases, pitfalls and failures for system under development. For well-developed ADS, it helps obtain trusts on decisions made and the system itself. Critical safety factors can be used to train human to gain a better driving skill for systems that outperform human and adopt the use of emerging technologies. Cornelissen {\it et al.} \cite{cornelissen2009} concluded that software systems ``must be sufficiently understood before it can be properly modified''.

To the best of our knowledge, there does not exist a framework that provides us a satisfactory understanding of ADS.
A number of studies have been attempted to ``understand'' the ADS system empirically only. Endsley \cite{endsley2017} reported the system behavior from end-users' perspective by sharing her empirical driving investigation with the Tesla Model S. The author noticed a number of undesirable {}behaviors but no systematic investigation was conducted. Teoh {\it et al.} \cite{teoh2017} analyzed records of the Google's AV after its road tests but was limited to statistical analyses and focused on occurred collisions only. 
On the other hand, the state-of-the-art MT-based frameworks such as DeepTest \cite{tian2018}, DeepRoad  \cite{zhang2018deeproad} and DeepHunter \cite{xie2018a} have been developed for testing only, but not for the purpose of understanding, and hence are unable to give us satisfactory knowledge about the system. That is to say, none of the previous frameworks was designed for a systematic understanding of the ADS.

In summary, although a comprehensive understanding of ADS is of great necessity, it remains unclear how to do it systematically. MT has been shown to be an effective technique to test the ADS, and will be demonstrated in this paper that it can be applied for the system understanding. In the next section, the proposed framework underpinned by MT technique will be described in more details.


\section{Proposed Framework}\label{section:framework}

\subsection{Motivation}

Let us start with the example illustrated in Figure~\ref{fig:SMG-example} and Figure~\ref{fig:SMG-process}. Suppose Autumn, one of the  well-known deep learning models in the Udacity challenge (\url{https://github.com/udacity/self-driving-car}) is used to predict steering angle ($\textsc{SA}_s$) for a given image (acting as the source image) (Figure~\ref{fig:SMG-example}a) by the target AV (i.e., the ego vehicle) whose ADS is under test. An MR is adopted to generate a follow-up image representing a scenario  where there is another car at a certain distance and of a certain color (Figure~\ref{fig:SMG-example}b) moving in front of the ego vehicle. The test outcome  of this MR requires that the predicted SA ($\textsc{SA}_f$) for the follow-up scenario must be smaller than a threshold $\kappa$, that is $|\textsc{SA}_s-\textsc{SA}_f|<\kappa$. It is similar to the criteria definition adopted in DeepTest \cite{tian2018}. 

To this end, the MR is applied to add a red car so that the source image and the follow-up image with the red car form an MG, namely MG1. The (normalized) SAs predicted by the Autumn model for the MG1 shown in Figure~\ref{fig:SMG-process}d are $\textsc{SA}_{1,s}=0.02$ and $\textsc{SA}_{1,f}=0.02$ (positive values indicates leftward/anticlockwise direction, negative ones represents rightward/clockwise tendency). With the MR violation condition being defined as of $\kappa=0.12$, no fault is detected (i.e., the MR satisfies $|0.02-0.02|<0.12$).

\vspace{0.5cm}
\begin{adjustbox}{center}
\includegraphics[scale=0.062]{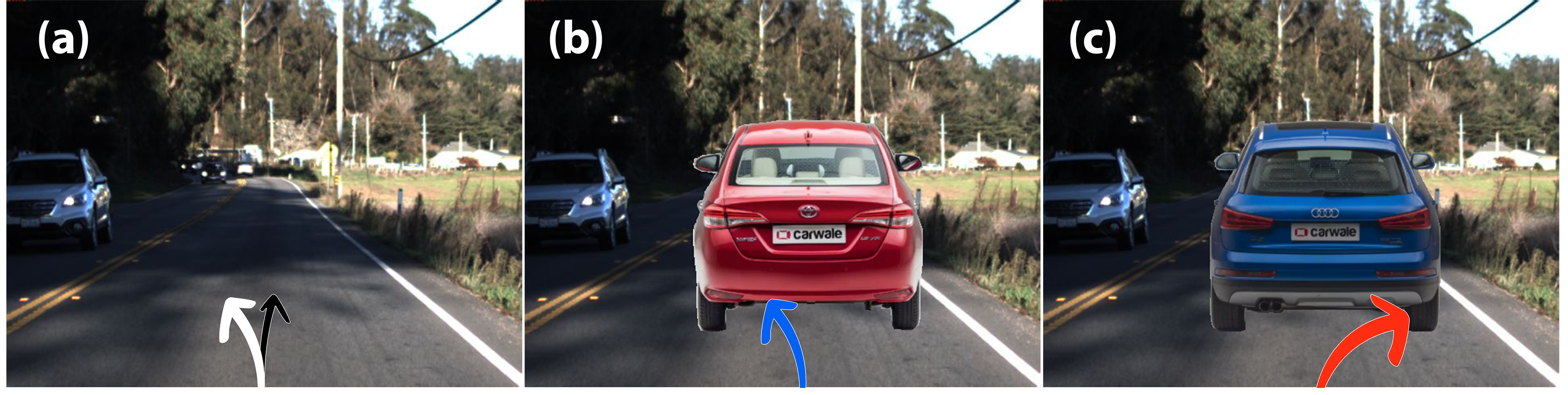}
\end{adjustbox}
\begin{adjustbox}{center,
caption={(a) Source image without car, (b) Follow-up image with red car inserted and (c) Follow-up image with blue car inserted at the same location as the red car in (b). White, red and blue arrows are the directions of SA predicted by the Autumn model for (a), (b) and (c) images, respectively. (Black arrows indicates the direction of SA from the ground truth when there is no leading car.) Specifically, the SAs (normalized to 1) predicted by the Autumn model for these images are (a) 0.02 leftward, (b) 0.02 leftward, and (c) 0.09 rightward.},
label={fig:SMG-example},
nofloat=figure}
\end{adjustbox}

\begin{algorithm}
\algnewcommand\algorithmicdef{\textbf{Data generation:}}
\algnewcommand\Definition{\item[\algorithmicdef]}
\algnewcommand\algorithmicexe{\textbf{System execution:}}
\algnewcommand\Execution{\item[\algorithmicexe]}
\newcommand{\Algrule}[1][.2pt]{\par\vskip.5\baselineskip\hrule height #1\par\vskip.5\baselineskip}
\caption{Outline of framework's algorithm.}
\label{algorithm}
\begin{algorithmic}
\Require \textit{\textbf{ inputs and user-defined functions}} \texttt{\\}
\textbullet ~Source dataset: $D$ \circled{1} \texttt{\\}
\textbullet ~Metamorphic relation: \textsc{MR} \circled{2} \texttt{\\}
\textbullet ~Set of configurations for SMGs: $C$ \circled{3}\texttt{\\} 
\textbullet ~User-defined function to generate an \textsc{MG} from an \textsc{MR}, a source datum $d$ and a configuration $c$: \Call{GenerateMG}{$\textsc{MR},d,c$} \circled{4} \texttt{\\}
\textbullet ~User-defined function to obtain an outcome from the target ADS: \Call{ExecuteADS}{$d$} \circled{5} \texttt{\\}
\textbullet ~User-defined function to determine undesirable behaviors from an MR, data $s,f$ and a configuration $c$: \Call{DetermineUB}{$\textsc{MR},s,f,c$} \circled{6} 
\Algrule
\Definition{\textit{\textbf{ given a metamorphic relation \textsc{MR}, the data $d_s$ and a configuration $C$}}}
\Function{GenerateSMG}{$\textsc{MR},d_s,C$}  \Comment{\circled{7}}
\State $G \gets \emptyset$ \Comment{A sequence of MG}
\For{$c \in C$}
\State $d_f \gets $ \Call{GenerateMG}{$\textsc{MR},d_s,c$}
\State $G \gets G \cup \{d_f\}$
\EndFor
\State \textbf{return} $G$
\EndFunction
\Algrule
\Execution{\textit{\textbf{given a metamorphic relation \textsc{MR}, a dataset $D$ and a configuration $C$}}}
\Function{Testing}{$\textsc{MR},D,C$}\Comment{\circled{8}}
\State $T \gets \emptyset$ \Comment{Set of testing outcomes}
\For{$d_s \in D$}
\State $S \gets \emptyset$ \Comment{Set of scenarios}
\State $U \gets \emptyset$ \Comment{Set of undesirable behaviors} 
\State $G \gets $ \Call{GenerateSMG}{$\textsc{MR},d_s,C$}
\State $o_s \gets $ \Call{ExecuteADS}{$d_s$}\Comment{Source SA}
\For{$(d_f,c) \in (G,C)$}
\State $o_f \gets $ \Call{ExecuteADS}{$d_f$}\Comment{Follow-ups SAs}
\State $u \gets $ \Call{DetermineUB}{$\textsc{MR},o_s,o_f,c$}
\State $S \gets S \cup \{(d_s,c)\}$
\State $U \gets U \cup \{u\}$
\EndFor
\State $T \gets T \cup \{(S,U)\}$
\EndFor
\State \textbf{return} $T$
\EndFunction
\end{algorithmic}
\end{algorithm}

\vspace{0.5cm}
\begin{adjustbox}{center}
\includegraphics[scale=0.038]{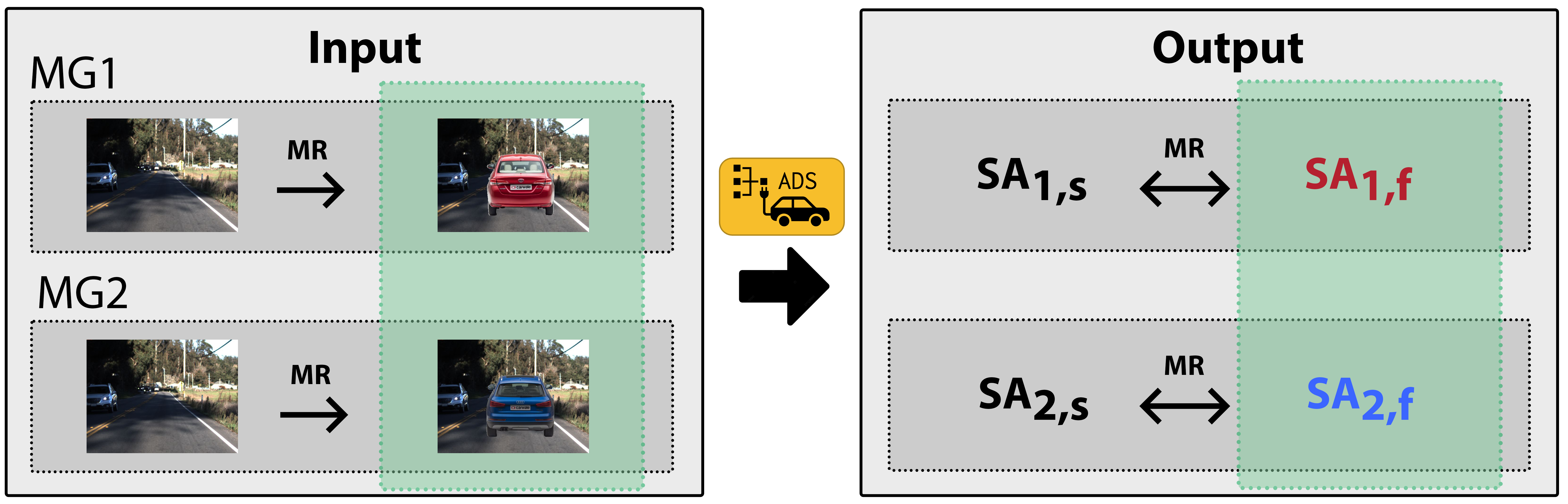}
\end{adjustbox}
\begin{adjustbox}{center,
caption={The illustration of process for generating and making further use of sequences of Metamorphic Groups (MG) with \textsc{Smart}. A source (original) image and a follow-up image generated by adding a leading car red using an MR form an MG, namely MG1. This MR is used to generate an image with a leading blue car in front, combining with the source image to construct the MG2. In the diagram, single-headed arrows indicate the generation of the follow-up image from the original one by the chosen MR and black double-headed arrows depict the MR validation from the SA outputs. Green areas denote the relationship between MG1 and MG2, and their corresponding outputs, which can be further utilized to detect the system's undesirableness.},
label={fig:SMG-process},
nofloat=figure}
\end{adjustbox}
\vspace{1cm}

In the following, one may be interested in knowing whether the color of the added car affects the decision by such an ADS system. For this reason, the MR is applied with a blue car (Figure~\ref{fig:SMG-example}c). 
The original image and a follow-up image with the blue car with exactly the same position and size form another MG, namely MG2. Again, the MR is satisfied with predicted SAs for MG2 ($\textsc{SA}_{2,s}=0.02$ and $\textsc{SA}_{2,f}=-0.09$, that is $|0.02-(-0.09)|<0.12$). Hence, no violation is detected, and one can argue that adding the blue car does not affect the decision because the system again only checks if the leading car is at a safe distance, and hence decides to steer right.

However, despite no violation is detected with these MGs, it is known that there is something undesirable with this system. The color of the leading car yields a significant change in the steering angle from left to right. It is argued that decision of a robust system should be consistent regardless of the color of the leading car. This undesirable behavior has only been revealed when MG1 and MG2 above are compared in a sequence. For this reason, a new framework is proposed in this paper to make use of sequences of MGs to conduct a systematic and in-depth testing and understanding of the ADS. 
In contrast to the existing methods such as DeepTest that separately check the MR violation of each individual MG (to determine erroneous behaviors), our new framework investigate the behaviors of a sequence of MGs in a collective way, as detailed in the following sections.


\subsection{Sequential Metamorphic Testing (\textsc{Smart}) Framework}

\subsubsection{The framework}

\begin{figure*}[!htbp]
	\centering
	\includegraphics[scale=0.175]{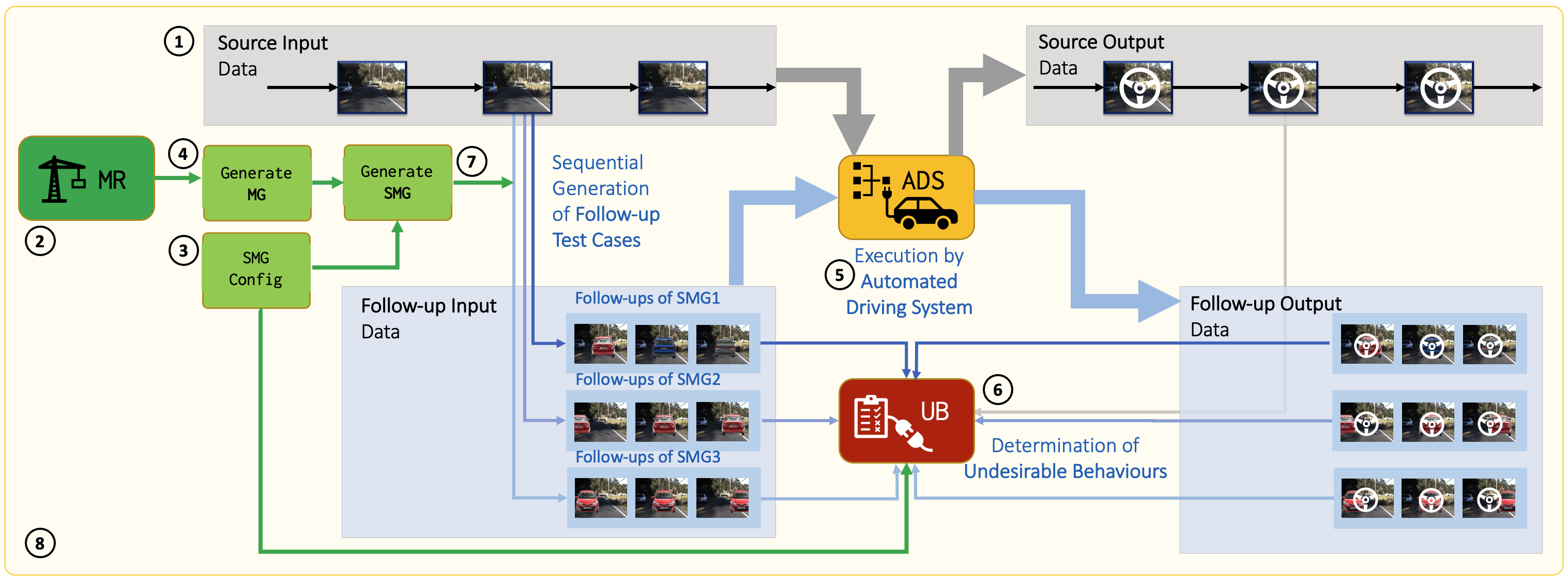}
	\caption{The \textsc{Smart} framework works without the need for ground truth data. The source input data can be used to generate different follow-up input data, forming sequences of metamorphic groups. They are executed by the ADS to produce corresponding outputs. Inputs and outputs of each SMG can be further examined to detect undesirable behaviors. Circled numbers are defined in Algorithm~\ref{algorithm}.}	
	\label{fig:framework}
\end{figure*}

The proposed Sequential MetAmoRphic Testing (\textsc{Smart}) framework consists of four main components, as shown in Figure~\ref{fig:framework}. First, it contains a set of MRs that facilitates the automatic generation of test cases and verification of their execution results. Second, \textsc{Smart} includes categories of MGs in a certain sequence generated from each MR. Third, quantitative metrics are defined to determine the degree of undesirable behaviors in the sequences of MGs. Last, the framework has a collection of necessary tools for test case generation, testing execution, and understanding of the ADS systems. The algorithm of \textsc{Smart} is presented in  Algorithm~\ref{algorithm}. Its source code is accessible at: \url{https://github.com/luuqh/smart}.

\subsubsection{Sequences of Metamorphic Groups}\label{SMG}

Given an MR and a set of source inputs, different MGs can be generated. In the \textsc{Smart}, MGs are designed and generated in a sequential way which have a rich relationship between them. Figure~\ref{fig:SMG-process} illustrates the progress for generating different MGs. While the proposed MR (Figure~\ref{fig:SMG-example}) may not be helpful in detecting any violation in the outputs, making use of the relationship between MGs may reveal undesirable behaviors of the system (Figure~\ref{fig:SMG-process}) as explained previously.
The guideline for designing a sequence of metamorphic groups (SMG) is given as follows:
\begin{enumerate} 
\item Determine the feature (or aspect) of the ADS for the understanding. For example, if it is necessary to understand the ADS' behaviors in response to the positions of car moving ahead, the positions can be considered as a feature.
\item Work out the range of values of each feature. For instance, the range of positions can be within the vision of AV. 
\item \label{step:smg} Divide the range into partitions. One naive way of partitioning is to divide it equally or proportionally. For example, the range can be divided into 400, 300, 200, 100 or zero pixels left or right away from the center of the car moving ahead. Note that different granularity levels of the partitioning refer to different precisions of the understanding.
\item Construct the SMG by creating one MG corresponding to each partition derived in Step \ref{step:smg}.
\end{enumerate}

{\color{revised}
What makes an SMG different from a regular set of MGs is the sequential way the MGs are implemented. By considering them in the sequential order, we can examine the variations of the undesirableness from a specific perspective instead of considering a whole set but without a particular order. For example, we can reveal at which position, scenario, or characteristics the system will respond undesirably.}

\subsection{Undesirable behavior metrics}

An {\it undesirable behavior} is a behavior that is considered abnormal by the system's user. In most cases, the metric \textsc{M} for an undesirable behavior can be defined by the common sense for a certain driving scenario. In the example above, when the system decides to steer leftward for the leading red car in MG1 and rightward for the leading blue car in MG2, this abnormal case might be referred to as an undesirable behavior, measuring by the difference between their corresponding steering angles. When the difference is larger than a certain threshold $\theta$, the Heaviside step function is adopted to determine an undesirable behavior, that is 
\begin{equation}
\begin{aligned}
	\textsc{U} = 
\begin{cases}
1, &\quad \textsc{M}>\theta\\
0, &\quad\text{otherwise}
\end{cases}	
\end{aligned}
\end{equation}
{where} $1$ is to assert that it is an undesirable behavior, and $0$ is to indicate otherwise or inclusiveness.
In general, it is up to the system's user to decide whether a behavior is undesirable or not by selecting the metric \textsc{M} and the threshold $\theta$. The metrics are not only to justify whether or not a scenario is an undesirable behavior, but also to measure the degree of undesirableness.

\section{Experiment}\label{section:experiment}

To demonstrate the effectiveness of \textsc{Smart}, experiments are conducted with deep learning models used to predict steering angles (SA). After the introduction of models to be tested, the metric of undesirable behaviors will be defined which will later help better understand the systems under test. The method to generate a specific sequence of MGs from selected MRs and the adoption of the metric for assessment are presented. 

\vspace*{-0.2cm}

\subsection{ADS systems under test}

To evaluate the proposed \textsc{Smart} framework, the experiments adopt deep learning models that are in the leaderboards of the Udacity Self-Driving Car Challenge 2 and have been used in earlier studies \cite{tian2018,zhang2018deeproad}. Among them, three models that have both source code and supporting data (trained weights, parameters) available, namely Chauffeur, Rambo and Autumn, are used. Given the set of input images, their output is the SA. Note that the SA decision to be tested in this study is only based on object detection from the traditional two-dimensional camera images without any depth information or perception that would be available from stereo cameras or LiDAR sensors. 

\begin{adjustbox}{center,
caption={Comparison between (a) Chauffeur model performance against the ground truth; and (b) Chauffeur model prediction without any car ahead during the trip and the one with the car. Dashed black line indicates the line of perfect fit where model prediction is exactly the same with the ground truth. 
Green color indicates higher density of points.},
label={fig:stats:chauffeur},
nofloat=figure,
vspace=\bigskipamount}
\includegraphics[width=4.4cm]{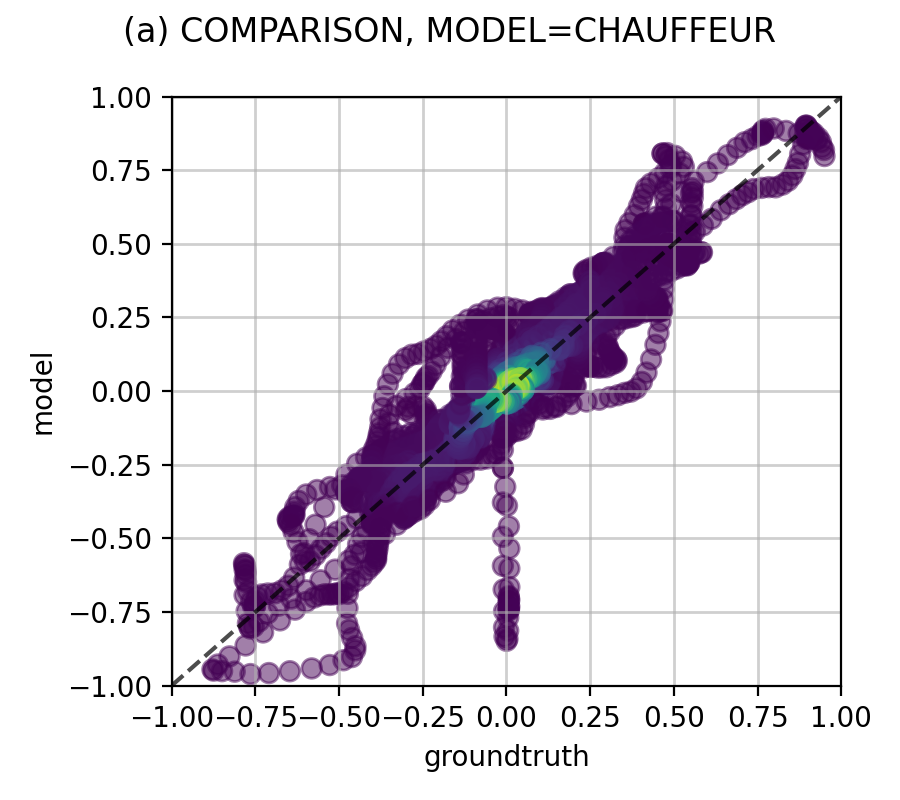}
\includegraphics[width=4.4cm]{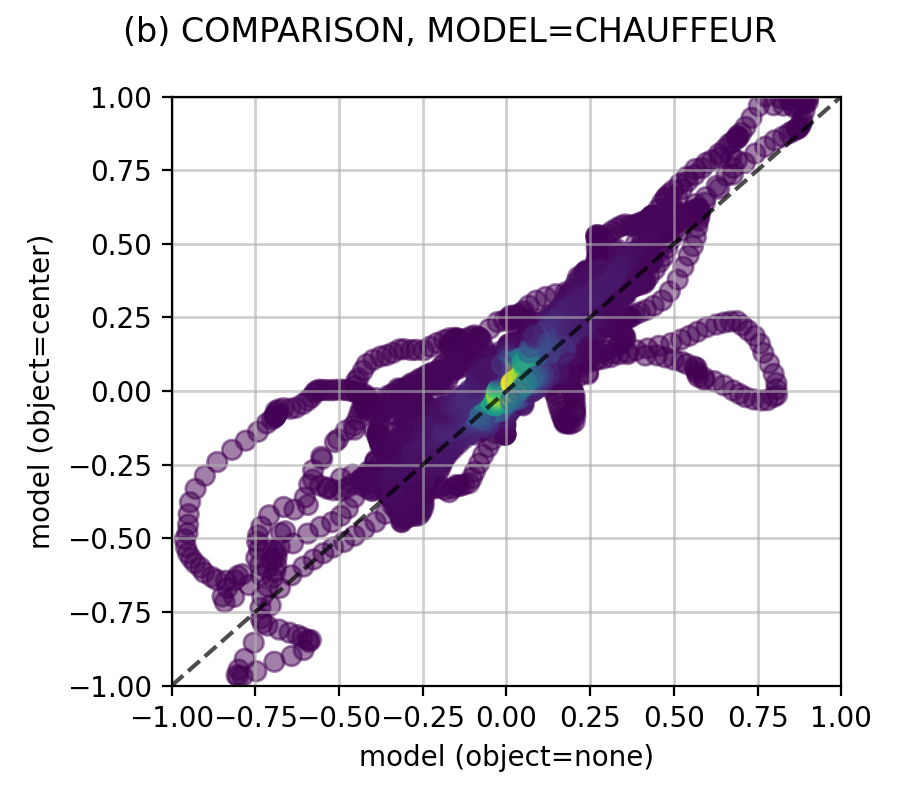}%
\end{adjustbox}

\begin{table}[!htbp]
\centering
\caption{Performance of selected ADS models}
\begin{center}
\label{table:models}
\begin{tabular}{l|llll}
Model & Chauffeur & Rambo & Autumn \\
\hline
mean absolute errors (MAE) & 0.055 & 0.074 & 0.050 \\
root mean square error (RMSE) & 0.092 & 0.101 & 0.098 \\
Pearson's correlation (CORR) & 0.910 & 0.885 & 0.884 \\
standard deviation (STDEV) & 0.091 & 0.100 & 0.098 \\
\end{tabular}
\end{center}
\end{table}


The overall architectures of both Chauffeur and Autumn models consist of a convolutional neural network (CNN) and a recurrent neural network (RNN). The CNN is used to extract dominant features from images, and Long Short-Term Memory (LSTM), a commonly used RNN, to predict the SA. Meanwhile, Rambo combines three different CNN streams for directly predicting the SA.

A Conda working environment that contains Python, TensorFlow, Keras, OpenCV and supporting libraries are built. To maintain the computability of Udacity models, the specific build (TensorFlow 1.15, Keras 1.2.2) is adopted. The experiments are carried out using the Swinburne Supercomputer (OZSTAR\footnote[1]{\url{https://supercomputing.swin.edu.au/ozstar/}}). Each of its compute nodes contains two Intel Gold 6140 18-core processors, two NVIDIA P100 12GB PCIe GPUs, 192 GB DDR4 RAM and 400 GB local SSD. 

The performance of these three models are presented in Table~\ref{table:models}. The mean absolute errors (MSE) of Chauffeur and Rambo were 0.055 and 0.074, respectively, which are similar to the results reported earlier  \cite{tian2018}. The performance of Autumn model is also in the same order of magnitude as other two models. The Pearson's correlation coefficients for all of them are high, in the range of 0.884 to 0.910. These data imply that these models are well configured and could be used for further investigation.

\subsection{Undesirable behavior metrics for steering angle}

The notion of {\it steering angle difference} (\textsc{AD}) is used to facilitate the evaluation of undesirable behavior metrics. \textsc{AD} is defined as the difference between a certain \textsc{SA} and a reference $\textsc{SA}^*$, that is,
\begin{equation}
\begin{aligned}
	\textsc{AD} = \textsc{SA} - \textsc{SA}^*
\end{aligned}
\end{equation}

To evaluate the model, follow-up images representing scenarios in a certain way are generated. The source image and its follow-up one form an MG. Assume the model prediction returns two values, $\textsc{SA}_s$ and $\textsc{SA}_f$. The $\textsc{SA}_s$ could be used as the reference to obtain the SA difference, and use it to validate the MR later. This difference is defined as
\begin{equation}
\begin{aligned}
	\textsc{AD} = \textsc{SA}_f - \textsc{SA}_s
\end{aligned}\label{eq:ad-mr}
\end{equation} 

Other frameworks, such as DeepTest, consider an MR to be violated when  $|\textsc{AD}|>\kappa$ for this MG. The proposed framework can further help unveil undesirable behaviors of the system, which works even in case there is no violation detected in individual MRs which are the basis for conventional MT-based frameworks. It is a result of harnessing MGs from a new perspective for determining the violation. For instance, given $\textsc{AD}_1$ and $\textsc{AD}_2$ computed from the outputs of MG1 and MG2 in the example (Figure~\ref{fig:SMG-process}). If the inputs are the same or have similar semantics, it is expected that $\textsc{AD}_1$ and $\textsc{AD}_2$ have a marginal difference. Otherwise, there is an undesirable behavior detected.

From an MR, multiple SMGs can be generated in a certain way. As a result, a set of \textsc{AD}s could be obtained, each of which is associated with an MG. For the ease of comparison, one of them can be selected as a reference, being denoted by $\textsc{AD}^*$. The change of \textsc{AD} against this baseline can be compared to determine the degree of undesirable behaviors. Quantitatively, the following metrics are used. 

The {\it unchange of steering} metric ($\textsc{U}_\textsc{N}$) is to capture the notion of no change in steering, that is, 
\begin{equation}
  \textsc{M}_\textsc{N} = 
 	\frac{1}{4}|\textsc{AD} - \textsc{AD}^*|
\end{equation}


The {\it rightward change of steering} metric ($\textsc{M}_\textsc{R}$) is to measure the rightward deviation of steering, that is, 
\begin{equation}
\begin{aligned}
  \textsc{M}_\textsc{R} = 
   	\begin{cases}
       \frac{1}{4}|\textsc{AD}-\textsc{AD}^*|,
	     &\quad\text{if } \textsc{AD} < \textsc{AD}^*\\
   	   0, 
   		 &\quad\text{otherwise}
	\end{cases}
\end{aligned}
\end{equation}

The {\it leftward change of steering} metric ($\textsc{M}_\textsc{L}$) is to measure the leftward deviation of steering, that is, 
\begin{equation}
\begin{aligned}
  \textsc{M}_\textsc{L} = 
   	\begin{cases}
       \frac{1}{4}|\textsc{AD}-\textsc{AD}^*|,
	     &\quad\text{if } \textsc{AD} > \textsc{AD}^*\\
   	   0, 
   		 &\quad\text{otherwise}
	\end{cases}
\end{aligned}
\end{equation}

\begin{table}
\centering
\caption{SMGs generated and used in this study.}
\begin{center}
\label{table:SMGs}
\begin{tabular}{l|l|l|c}
\hline
Name & Description & Configuration for generating & Reference\\
 & &  follow-up photos & MG\\
\hline
SMG1 & Forward & left 400 pixels, left 300 pixels, & center \\
&  moving car & ~ left 200 pixels, left 100 pixels, &  \\
&  & ~ center, right 100 pixels, &\\
&  & ~ right 200 pixels, right 300 &\\
&  & ~  pixels, right 400 pixels & \\
SMG2 & Approaching & left 400 pixels, left 300 pixels, & center \\
& car & ~ left 200 pixels, left 100 pixels, &  \\
&  & ~ center, right 100 pixels, &\\
&  & ~ right 200 pixels, right 300 &\\
&  & ~  pixels, right 400 pixels & \\
SMG3 & Different & red, blue, grey & red \\
 & leading car & \\
SMG4 & Combined & car, car+snow:0.2, car+snow:0.4, \\
 & car \& weather & car+snow:0.6, car:snow:0.8,& car \\
& & car+snow:1.0 & \\ 
\hline
\end{tabular}
\end{center}
\end{table}

\begin{adjustbox}{center}
	\includegraphics[scale=0.12]{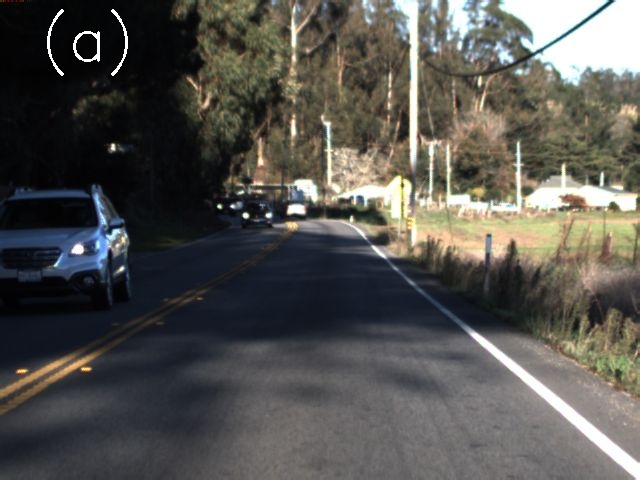}
	\includegraphics[scale=0.12]{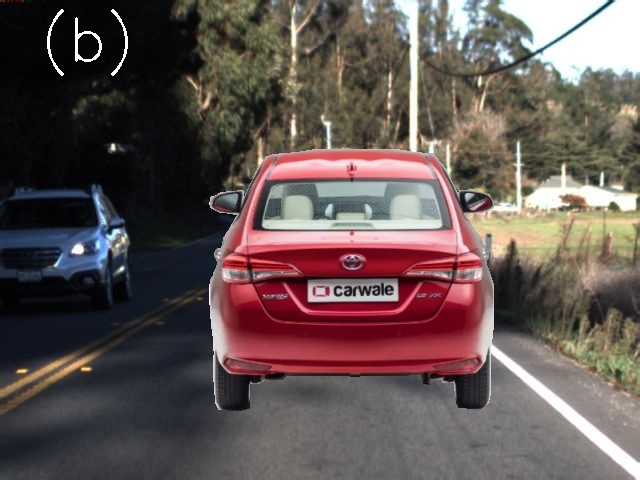}
	\includegraphics[scale=0.12]{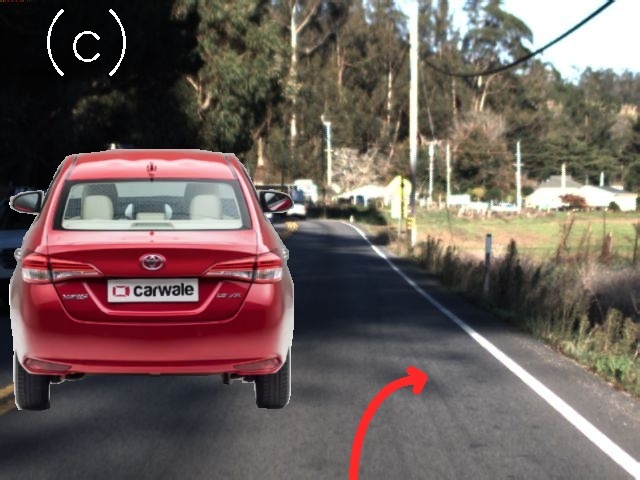}
\end{adjustbox}
\begin{adjustbox}{center}
	\includegraphics[scale=0.12]{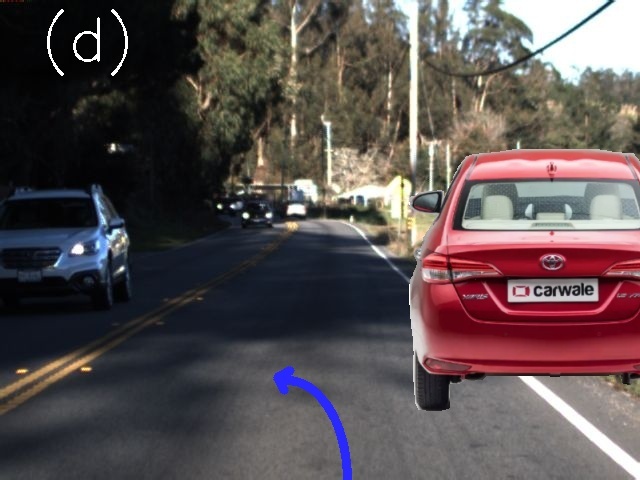}
	\includegraphics[scale=0.12]{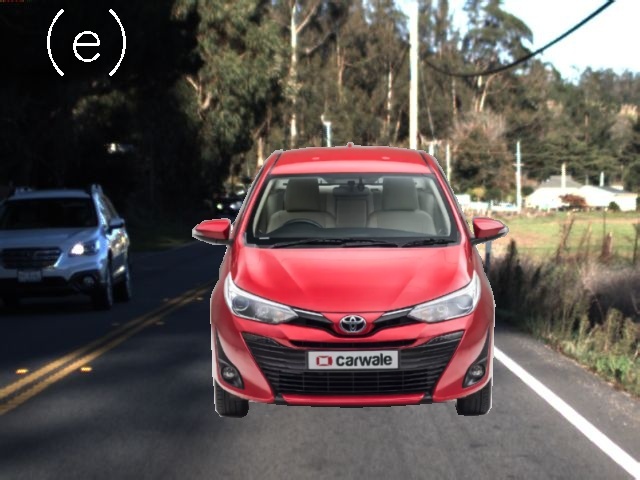}
	\includegraphics[scale=0.12]{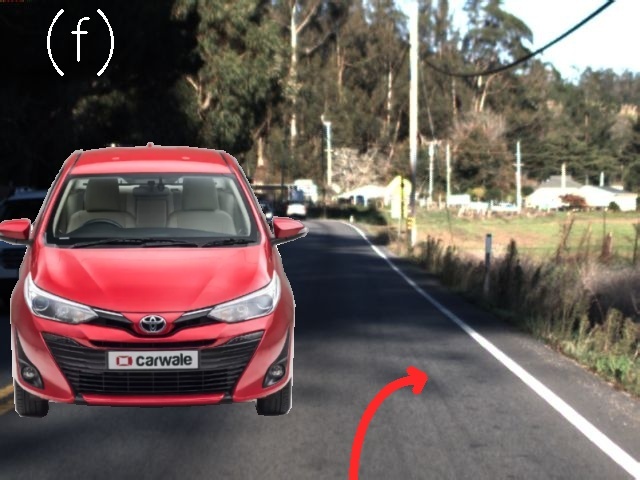}
\end{adjustbox}
\begin{adjustbox}{center}
	\includegraphics[scale=0.12]{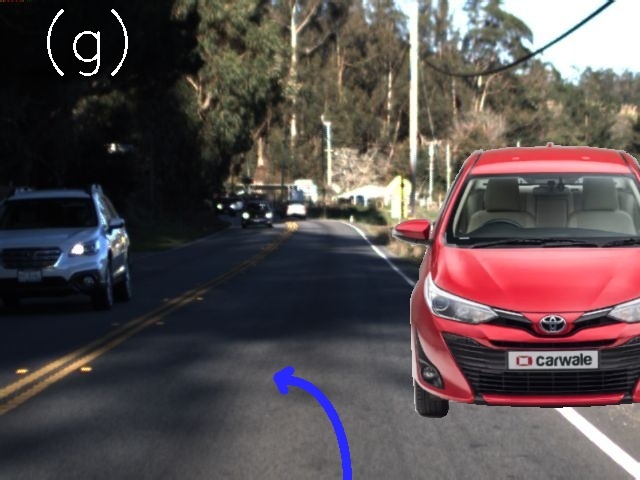}
	\includegraphics[scale=0.12]{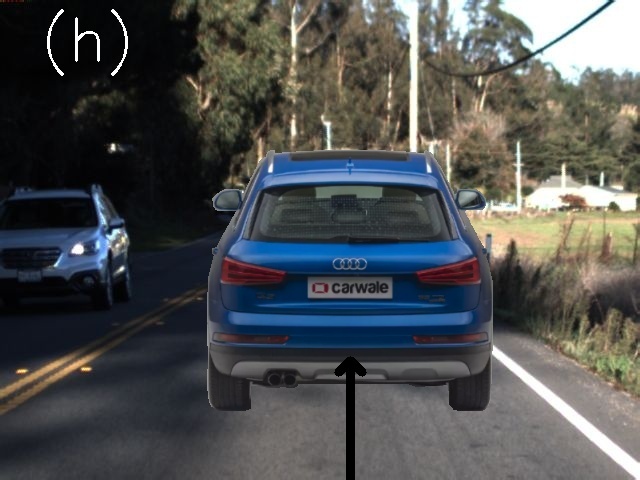}
	\includegraphics[scale=0.12]{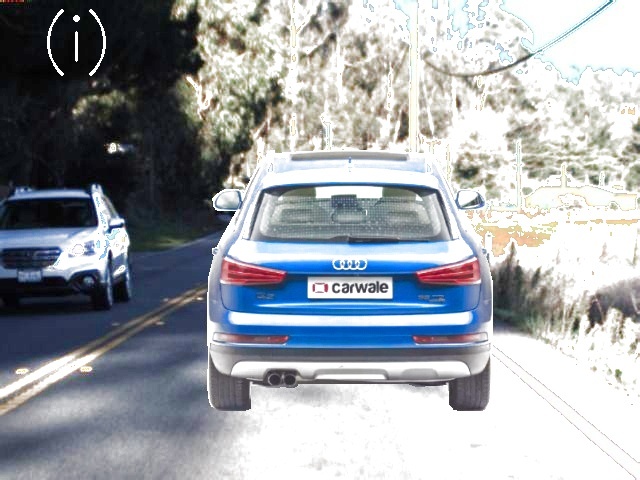}
\end{adjustbox}
\begin{adjustbox}{center,
caption={Snapshot of images from MGs: (a) Source image; and follow-up images with (b) forward (red) car, (c) forward left-side car, (d) right-side car, (e) facing car, (f)  facing left-side car, (g) facing right-side car, (h) forward blue car, and (i) forward car under a snowy weather. The blue and red arrows shows expected leftward and rightward steering angles \textsc{AD}, respectively, in relative to $\textsc{AD}^*$.},
label={fig:underdesirable-examples},
nofloat=figure}
\end{adjustbox}
\vspace{\bigskipamount}


It is up to the users to define which metric should be adopted in a certain driving scenario. In this study, only one metric is used at a time for a scenario. Since $\textsc{AD}$ and $\textsc{AD}^*$ are in the range $[-2,2]$ for normalized SAs (between $-1$ and $1$ {\cite{tian2018}}), the metrics (${\textsc{M}}_\textsc{N}$, ${\textsc{M}}_\textsc{R}$ and ${\textsc{M}}_\textsc{L}$) are in the range of $[0,1]$ with the adjustment of the fraction $1/4$. Here, 1 and 0 indicate the highest and the lowest degrees, respectively, associated with the (chosen undesirable) behaviors. 

The steering angle for the source image at a certain timeframe is used as a reference $\textsc{SA}^*$ . The reference $\textsc{AD}^*$  varies, subject to the MGs. It is listed in the last column of Table~\ref{table:SMGs}.

\subsection{Sequences of Metamorphic Groups}

Given an MR to generate a car image at a certain distance ahead and the source input $I_s$ consisting of a source image (Figure~\ref{fig:underdesirable-examples}a).  The follow-up image $I_f$ (Figure~\ref{fig:underdesirable-examples}b) is generated using the MR, forming an MG ($I_s,I_f$). $\textsc{AD}$ is obtained from the outputs of ADS after its execution. From the same source image, the follow-up images are generated in a horizontal or a certain sequence (Figure~\ref{fig:underdesirable-examples}), obtaining a set of $\textsc{AD}$s. Among them, $\textsc{AD}^*$ is selected for each SMG as shown in Table~\ref{table:SMGs}, and then defined the undesirable behaviors. The details are as follows. 

\subsubsection{SMG1: Forward moving car}

The follow-up input $I_f$ is an image being generated by adding a forward moving car to the source image, either on the left (Figure~\ref{fig:underdesirable-examples}c), in the middle (Figure~\ref{fig:underdesirable-examples}b) or on the right (Figure~\ref{fig:underdesirable-examples}d).

\begin{itemize}
\item The metric $\textsc{M}_\textsc{L}$ is applied to measure the undesirable behavior of the ADS when it steers leftward (anticlockwise) in response to the forward car on the left (Figure~\ref{fig:underdesirable-examples}c).
\item The metric $\textsc{M}_\textsc{R}$ is applied to measure the undesirable behavior of the ADS when it steers rightward (clockwise) in response to the forward moving car on the right (Figure~\ref{fig:underdesirable-examples}d).
\end{itemize}

\subsubsection{SMG2: Approaching car}

The follow-up input $I_f$ is an image being generated by adding an approaching car in front of the ego car, either on the left (Figure~\ref{fig:underdesirable-examples}e), in the middle (Figure~\ref{fig:underdesirable-examples}f) or on the right (Figure~\ref{fig:underdesirable-examples}g).

\begin{itemize}
\item The metric $\textsc{M}_\textsc{L}$ is applied to measure the undesirable behavior of the ADS when it steers leftward (anticlockwise) in response to the approaching car on the left (Figure~\ref{fig:underdesirable-examples}f).
\item The metric $\textsc{M}_\textsc{R}$ is applied to measure the undesirable behavior of the ADS when it steers rightward (clockwise) in response to the approaching car on the right (Figure~\ref{fig:underdesirable-examples}g).
\end{itemize}

\subsubsection{SMG3: Different leading car}

The follow-up input $I_f$ is an image being generated by adding a forward moving car, either in red (Figure~\ref{fig:underdesirable-examples}b),  blue (Figure~\ref{fig:underdesirable-examples}h) or grey color.

\begin{itemize}
\item The metric ${\textsc{M}}_\textsc{N}$ is applied to measure the undesirable behavior of the ADS when it steers either leftward (anticlockwise) or rightward (clockwise) in response to the leading car with different colors.
\end{itemize}

\subsubsection{SMG4: Combined moving car and weather}
The follow-up input $I_f$ is an image being generated by adding a car and transforming the scene into a snowy weather with different intensities  (Figure~\ref{fig:underdesirable-examples}i).

\begin{itemize}
\item The metric ${\textsc{M}}_\textsc{N}$ is applied to measure the undesirable behavior of the ADS when it steers differently against the scenario with a snowy weather.
\end{itemize}

For a complete list of experiments, please refer to the supplementary materials and the framework’s public repository  (\url{https://github.com/luuqh/smart/}). 

Let us look at an example. In the given dataset, the ego car is driving in a single-lane road. In this scenario, the behavior of an artificial car would be undesirable and hazardous if it is on the centre of a dividing line. In response to that, the ego car may decide to go straight if it perceives that the distance is safe to do so, or to steer left to avoid a potential collision. However, if it steers to the right, its tendency is to move outside of the current lane and into the roadside areas, which is an undesirable behavior.

\vspace{-0.3cm}
\subsection{Test case generations and evaluations}

The datasets are obtained from the Udacity where images are extracted from cameras placed behind the wind-shield of the ego car, and SA of the human driver was recorded. Three models are pre-trained with the CH\_001 dataset consisting of 101,397 photos and labelled SA, covering a wide range of driving conditions. This study adopts the CH\_002 dataset, the same set being used for evaluating the performance of models participating in the competition, as the source inputs. It consists of 5,615 photos captured by the center camera in a resolution of $640 \times 480$ pixels. 
To generate the sequential follow-up inputs, a package based on the OpenCV library is developed, which enables rescaling and inserting a chosen object to any location in the original photo. It is used to generate four SMGs as in Table~\ref{table:SMGs}. To examine the feasibility and robustness of the generated datasets, all generated images have been carefully inspected one by one to ensure that they are 
realistic. For example, when the image shown in Figure~\ref{fig:underdesirable-examples}c was generated, it was observed that the added artificial car actually blocks the view the car in source image. After examining these two cars' different sizes and positions, it could be confirmed that they were not overlapped, that is, it is still realistic and feasible to have the artificial car in Figure~\ref{fig:underdesirable-examples}c, which happens to make part of the car in source image invisible to the ego car.
Two thresholds are used to evaluate the models. With the low threshold $\theta=0$, it is aimed to reveal all possible undesirable behaviors. For the high threshold $\theta=0.02$ (which is the same order as of $25\%$ of standard deviations as in Table~\ref{table:models}), the main focus is on severe undesirable behaviors only.

\vspace{0.2cm}
\section{Result and Discussion}\label{section:results}

\subsection{The effectiveness of framework in uncovering undesirable behaviors of the ADS}

The challenge is that in general, there is no way to determine whether or not the decision made by model is correct or wrong, which implies whether or not there is an issue in the ADS system. A statistical comparison (Figure~\ref{fig:stats:chauffeur}a) between the model prediction (Figure~\ref{fig:SMG-process}a, white arrow) and the ground truth (Figure~\ref{fig:SMG-process}a, black arrow) may not be helpful because the difference is often attributed to model performance. A similar statistical analysis shown in Figure~\ref{fig:stats:chauffeur}b may not be affirmative to determine whether there is a fault in the ADS. Alternatively, one may use a scenario  with a car moving forward in front of the ego car (Figure~\ref{fig:underdesirable-examples}b) to see how the ADS responses against the original scenario where there is no car ahead (Figure~\ref{fig:underdesirable-examples}a).
One may argue that the SA difference between the scenario with a car ahead and the one without it can be referred to as the impact of the object.

\begin{figure*}[!htb]
	\centering
	\includegraphics[scale=0.42]{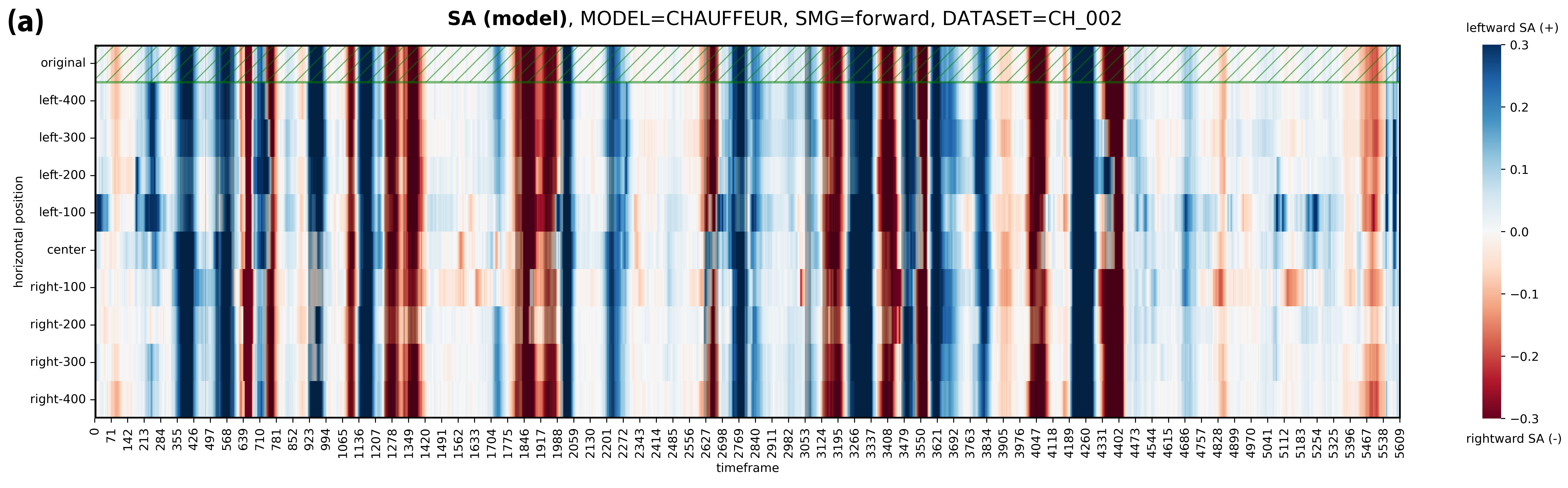}
	\includegraphics[scale=0.42]{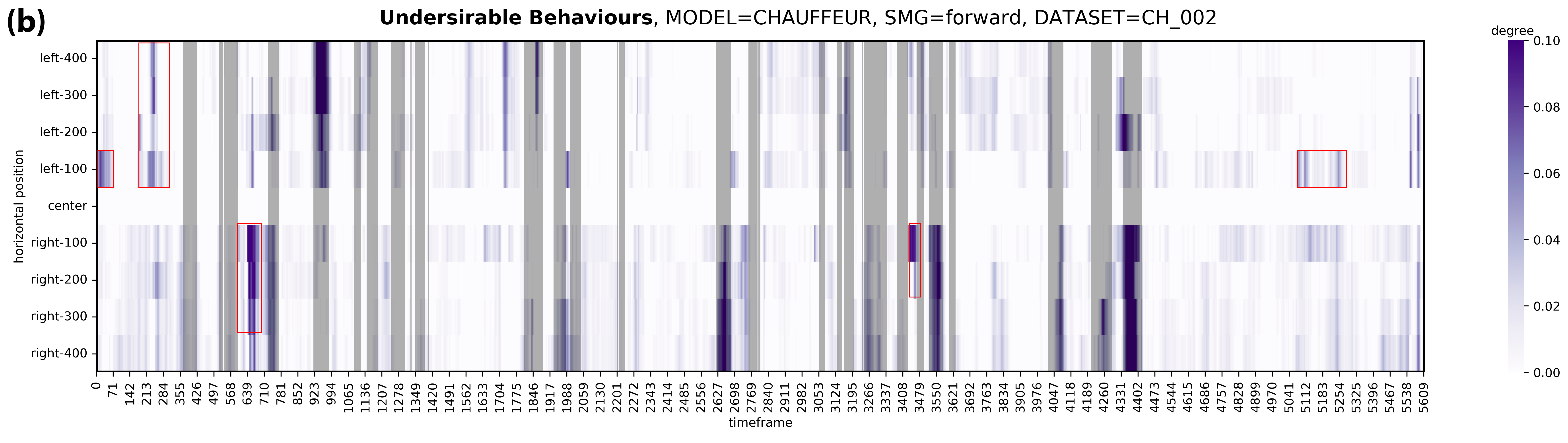}	
	\caption{(a) Heatmap of raw steering angles of Chauffeur model during the trip for different positions of a car moving forward (SMG1). Blue color represents leftward SA whilst red color shows rightward SA. Green hatch areas highlight SAs for original images. (b) Heatmap of undesirable behaviors of Chauffeur model during the trip as from the examination of SMG1. Purple color represents the high degree of undesirable behaviors. Grey areas highlight (24.1\%) road segments that are curved (defined by $\textsc{SA}>0.2$ from ground truth data). Red rectangles highlight some notable regions with high degree of undesirable behaviors.
	\label{fig:sa:chauffeur}}
\vspace{-0.5cm}
\end{figure*}

With the \textsc{Smart} framework, the behaviors of the systems can be examined in a fine-grained scale. Figure~\ref{fig:sa:chauffeur}a depicts a heatmap representing the ``raw'' predicted SA of Chauffeur model for different positions of a car moving forward. From the heatmap, scenarios and time where the ADS decides to steer leftward, manoeuvre to the right or go straight (presenting by blue, red and grey colors in Figure~\ref{fig:sa:chauffeur}a, respectively) can be compared. Furthermore, by adopting the reference MG in constructing SMGs (Table~\ref{table:SMGs}), the relative difference between MGs within SMGs is obtainable. The main product of the framework is the heatmap displaying the degree of undesirable behaviors (Figure~\ref{fig:sa:chauffeur}b). From the map, one could locate a subset (by means of timeframe) of the dataset that has a higher degree of undesirable behaviors (highlighted by red rectangles in Figure~\ref{fig:sa:chauffeur}b), which is referred to as the explanation-by-examples \cite{yang2021}. On the other hand, it is possible to identify which feature in the image associated with the car that has the higher influence on the undesirable behaviors, what is referred to as explanation-by-features \cite{yang2021}. For example, the Chauffeur model has more issues when the moving car is at the rightmost position (right-400 in Figure~\ref{fig:sa:chauffeur}b) other than it is closer to the center (right-100 in Figure~\ref{fig:sa:chauffeur}b).


\begin{table}
\centering
\caption{Undesirable behaviors of different models (straight roads only).}
\begin{center}
\label{table:ub}
\begin{tabular}{l|rr|rr|rr}
 \hline
  Model & \multicolumn{2}{c}{Chauffeur} &
  \multicolumn{2}{c}{Rambo} &
  \multicolumn{2}{c}{Autumn} \\
 \hline
Threshold   & $0$   & $0.02$   & $0$   & $0.02$ & $0$   & $0.02$ \\ 
{\cellcolor{blue!15}SMG1} &
{\cellcolor{blue!15}43.2\%} &
{\cellcolor{blue!15}8.0\%} &
{\cellcolor{blue!15}45.0\%} &
{\cellcolor{blue!15}4.9\%} &
{\cellcolor{blue!15}45.4\%} &
{\cellcolor{blue!15}5.2\%}\\
 left-400   & 30.3\%     & 4.4\%   & 45.3\%   & 1.9\%   & 39.3\%    & 3.5\%   \\
 left-300   & 34.6\%     & 6.4\%   & 54.5\%    & 5.9\%  & 49.8\%    & 8.4\%   \\
 left-200   & 33.7\%     & 6.5\%    & 62.9\%     & 13.4\% & 52.5\%    & 15.1\%   \\
 left-100   & 39.1\%     & 8.3\%  & 59.2\%    & 2.9\%  & 58.4\%     & 4.2\%   \\
{\cellcolor{green!10}left} &
{\cellcolor{green!10}34.4\%} &
{\cellcolor{green!10}6.4\%} &
{\cellcolor{green!10}55.5\%} &
{\cellcolor{green!10}6.0\%} &
{\cellcolor{green!10}50.0\%} &
{\cellcolor{green!10}7.8\%}\\
 right-100   & 55.0\%   & 11.7\%   & 42.9\%     & 3.6\%    & 53.1\%     & 3.9\% \\
 right-200   & 47.5\%    & 8.1\%    & 40.6\%     & 5.4\%   & 46.5\%    & 3.1\%  \\
 right-300   & 50.7\%     & 8.0\%   & 19.5\%   & 4.3\%   & 34.4\%    & 1.8\%  \\
 right-400   & 55.0\%    & 10.9\%  & 35.2\%     & 2.0\% & 28.8\%     & 1.5\%  \\ 
{\cellcolor{green!10}right} &
{\cellcolor{green!10}52.1\%} &
{\cellcolor{green!10}9.7\%} &
{\cellcolor{green!10}34.6\%} &
{\cellcolor{green!10}3.8\%} &
{\cellcolor{green!10}40.7\%} &
{\cellcolor{green!10}2.6\%}\\
{\cellcolor{blue!15}SMG2} &
{\cellcolor{blue!15}35.5\%} &
{\cellcolor{blue!15}13.2\%} &
{\cellcolor{blue!15}46.5\%} &
{\cellcolor{blue!15}5.2\%} &
{\cellcolor{blue!15}47.8\%} &
{\cellcolor{blue!15}6.6\%}\\
 left-400   & 39.9\%   & 11.5\%   & 44.7\%   & 2.1\%  & 40.9\%    & 3.6\%   \\
 left-300   & 35.6\%   & 13.6\%   & 54.8\%    & 7.1\%  & 48.0\%    & 5.9\%   \\
 left-200   & 30.7\%  & 12.4\%   & 62.4\%   & 13.4\%   & 58.7\%   & 21.9\%   \\
 left-100   & 41.6\%    & 15.3\%   & 58.1\%    & 3.3\%   & 66.4\%    & 5.7\%   \\
{\cellcolor{green!10}left} &
{\cellcolor{green!10}37.0\%} &
{\cellcolor{green!10}13.2\%} &
{\cellcolor{green!10}55.0\%} &
{\cellcolor{green!10}6.5\%} &
{\cellcolor{green!10}53.5\%} &
{\cellcolor{green!10}9.3\%}\\ 
 right-100   & 30.3\%    & 8.1\%   & 38.4\%    & 3.1\%  & 55.5\%   & 4.5\%   \\
 right-200   & 28.1\%   & 12.9\%   & 40.3\%    & 5.7\%  & 47.7\%   & 4.3\%   \\
 right-300   & 36.0\%    & 15.4\%   & 39.6\%   & 4.7\%  & 33.7\%    & 4.0\%   \\
 right-400   & 41.5\%    & 16.1\%   & 33.7\%    & 1.8\%   & 31.2\%  & 2.7\%   \\
{\cellcolor{green!10}right} &
{\cellcolor{green!10}34.0\%} &
{\cellcolor{green!10}13.1\%} &
{\cellcolor{green!10}38.0\%} &
{\cellcolor{green!10}3.8\%} &
{\cellcolor{green!10}42.0\%} &
{\cellcolor{green!10}3.9\%}\\ 
{\cellcolor{blue!15}SMG3} &
{\cellcolor{blue!15}75.9\%} &
{\cellcolor{blue!15}18.9\%} &
{\cellcolor{blue!15}75.8\%} &
{\cellcolor{blue!15}0.1\%} &
{\cellcolor{blue!15}65.6\%} &
{\cellcolor{blue!15}0.6\%}\\
 car-blue   & 75.9\%     & 26.6\%   & 75.8\%   & 0.0\%   & 65.4\%    & 0.5\%   \\
 {car-grey}   & 75.9\%   & 11.2\%    & 75.8\%     & 0.2\%   & 65.7\%     & 0.7\%   \\
{\cellcolor{blue!15}SMG4} &    {\cellcolor{blue!15}75.9\%} &    {\cellcolor{blue!15}35.8\%} &    
{\cellcolor{blue!15}75.8\%} &    
{\cellcolor{blue!15}15.0\%} &    
{\cellcolor{blue!15}70.8\%} &    {\cellcolor{blue!15}1.7\%}\\
car+snow:0.2 &    75.9\% &    38.6\%&    75.8\% &    15.8\% &    68.1\% &    1.2\% \\
car+snow:0.4 &    75.9\% &    38.8\%&    75.8\% &    15.0\% &    74.3\% &    2.2\% \\
car+snow:0.6 &    75.9\% &    33.7\%&    75.8\% &    13.8\% &    74.3\% &    2.1\% \\
car+snow:0.8 &    75.9\% &    33.2\%&    75.8\% &    14.4\% &    68.5\% &    1.2\% \\
car+snow:1.0 &    75.9\% &    34.6\%&    75.8\% &    15.8\% &    69.0\% &    1.7\% \\
\hline
\end{tabular}
\end{center}
\end{table}

The sequential combination of SMGs helps better understand the undesirableness of ADS.  Table~\ref{table:ub} depicts the summary of undesirable behaviors revealed for Chauffeur model in response to the car. The figure shows that our \textsc{Smart} framework could detect the undesirableness in the model for different scenarios corresponding to different positions of the car moving forward (SMG1), either in the left or in the right. The degrees of undesirable behaviors are surprisingly high, ranging from 30.3\% to 55.0\% {(with an average of 43.2\%)} for the low threshold ($\theta=0$) and from 4.4\%--11.7\% {(with an average of 8.0\%)}  for the high threshold ($\theta=0.02$) (Table~\ref{table:ub}). In other words, from 5,615 original photos, total numbers of 19,420 and 3,610 scenarios associated with low and high thresholds, respectively, of undesirable behaviors in the Chauffeur model are identifiable just by examining the SMG1. These high percentages indicate that the model is much more problematic than one would expect, highlighting the effectiveness of our framework.
 
Similar high effectiveness of \textsc{Smart} in detecting undesirable behaviors of the model could be observed in other SMGs. For example, SMG2 contains different scenarios, each corresponding to a position of the approaching car. The degrees of undesirable behaviors of SMG2 are slightly lower than that of SMG1 (Table~\ref{table:ub}) for the low threshold, ranging from 30.3\% to 41.6\%. However, more severe issues are found in the facing car for the high threshold, with values ranging 4.4\%--11.7\% (Table~\ref{table:ub}) for SMG2. 
The highest effectiveness found among all SMGs examined is with the SMG3 and SMG4. While replacing the red car by either blue or grey car, the degrees of undesirable behaviors for the low threshold are as high as 75.9\% (Table~\ref{table:ub}). 
For the high threshold \textsc{Smart} reveals that 11.2\%--38.8\% (Table~\ref{table:ub}) of all decisions made by the Chauffeur model are undesirable. 

In brief, \textsc{Smart} is highly effective in revealing undesirable behaviors. It helps address 30.3\%--75.9\% of undesirable scenarios for the low threshold and 4.4\%--38.8\% for the high threshold with four SMGs, equivalent to several thousands of undesirable scenarios.
 

\subsection{Applicability of the framework in distinguishing different ADS systems}

The performance of \textsc{Smart} on three different ADS models, namely Chauffeur, Rambo and Autumn, are compared. 
First, it is found that \textsc{Smart} not only is highly effective in revealing undesirable behaviors with Chauffeur model, but also performs well for Rambo and Autumn models. In response to the car moving forward (SMG1), the Rambo has a ratio of 19.5\%--62.9\% undesirableness among all scenarios tested (Table~\ref{table:ub}). It is also indicated that 28.8\%--58.4\% predictions using the low threshold by Autumn model are undesirable (Table~\ref{table:ub}). They are in the same order of magnitude as of Chauffeur model (30.3\%--55.0\%) (Table~\ref{table:ub}). For the high threshold, the degrees of undesirable behaviors for Rambo and Autumn models are in the ranges of 1.9\%--13.4\% and 1.5\%--15.1\%, respectively (Table~\ref{table:ub}), which are comparable to the Chauffeur model. Similar results are observed in other SMGs. For the SMG2 (related to the approaching car), \textsc{Smart} reveals the same order of undesirable behaviors in the Rambo and the Autumn models in comparison with the Chauffeur model (Table~\ref{table:ub}). For the SMG3 (on having an alternative appearance for the leading car) and SMG4, \textsc{Smart} can reveal a high percentage of issues (Table~\ref{table:ub}). These results demonstrate that the effectiveness of \textsc{Smart} is consistent across different models.

Having said that, \textsc{Smart} can distinguish fine details in the responses of different models. For example, the Chauffeur model has more severe undesirable behaviors (that is, having a larger number of undesirable behaviors at the high threshold in comparison with other models) for the scenario with the car moving forward (SMG1) and positioned 100 pixels rightward (Table~\ref{table:ub}) compared to other positions. In contrast, both Rambo and Autumn models are prone to issues in which the car is moving forward and positioned 200 pixels leftward (Table~\ref{table:ub}). For the SMG2 (Table~\ref{table:ub}), both Rambo and Autumn have the highest percentage of undesirableness for the high threshold at the car positioned 200 pixels leftward, being 13.4\% and 21.9\%, respectively. In comparison, the Chauffeur model has a high percentage (>10\%) for most MGs. For the SMG3 (Table~\ref{table:ub}), only the Chauffeur model has a high degree of undesirable behaviors for both cases (11.2\%--26.6\%). Other two models have a low percentage (<1\%) of undesirable behaviors. For the SMG4, Chauffeur is still the least robust model, whilst the best one is Autumn model, especially for the high threshold (Table~\ref{table:ub}).

In a word, \textsc{Smart} remains highly effective in revealing undesirable behaviors of different models and can be used to distinguish them in a fine-grained scale.

\section{Insights into ADS' Decision-Making Process}\label{section:insights}

The \textsc{Smart} framework helps reveal different critical factors and features affecting the ADS. As a result, it may help gain deeper insights into their decisions, some of which have never been provided by the other traditional MT-based techniques. Particularly, the proposed \textsc{Smart} framework helps uncover the (un)favorable factors, the extent of responses to hazardous scenarios, the robustness and consistency and new critical features, as discussed in the following.

\vspace*{-0.4cm}
\subsection{Determination of (un)favorable factors of each ADS}

The framework enables to, for example, examine whether or not there is any (un)favorable factor in the decision for scenarios related to different positions of the car moving ahead for each ADS model. It has been found that in responding to the car moving forward (SMG1), the Chauffeur model has more undesirable behaviors if the car ahead is in the right side of the ego car other than in the left side (Table~\ref{table:ub}). In details, the average of undesirable behaviors of the left side-scenarios is 34.4\%, which is smaller than the right-side value as of 52.1\% (Table~\ref{table:ub}). In contrast, both the Autumn and the Rambo models have more undesirable behaviors for the forward moving car on the right side whose averages values are 20.9\% and 9.3\%, respectively (for the low threshold, Table~\ref{table:ub}). In other words, the Chauffeur model tends to have more issues on the forward moving car on the right, compared to the Rambo and Autumn models which contain more undesirableness on the left. 

\vspace*{-0.2cm}
\subsection{Identifying the extent of response to hazardous scenarios}

It is unknown whether or not specific hazardous scenarios will cause undesirable behaviors in the ADS and, if it is the case, to what extent. It is the aspect that has not been investigated in existing MT-based frameworks. In our framework, for instance, a comparison of results from the sequence of scenarios corresponding to different positions of the approaching car can provide such information. In particular, it is found that all three models have the same severe problem with the scenario for the car moving forward at the position of 200 pixels leftward (Table~\ref{table:ub}). Similar undesirable issues are observed with the MGs related to the scenario for the approaching car at the position of 200 pixels leftward (Table~\ref{table:ub}). This position is associated with the hazardous scenario where the car is on the continuous dividing lines between two opposite lanes. All models have struggled to make good decisions for this scenario. An obvious suggestion is that the future ADS model should be trained to cope with such unusual but risky scenarios. 

\vspace*{-0.2cm}
\subsection{Inspecting the robustness against a certain feature}

It is important to ensure that the ADS decisions are robust and consistent against a feature, which has not been paid enough attention in existing MT-based frameworks. For instance, the robustness of a model against the change of appearance could be examined by our frameworks. It has been found that the decisions made by both the Rambo and the Autumn models are quite consistent. There are undesirable behaviors for the low threshold, but only a low percentage ($<1$\%) of severe undesirable behaviors found in both models (Table~\ref{table:ub}). The result indicates that these two models are quite robust against different appearance of the car. Meanwhile, the decision of Chauffeur is sensitive to the colors of the car with more severe undesirable behaviors (11.2\%--26.6\% in Table~\ref{table:ub}). In a word, the characteristics of object are a critical factor affecting the robustness of the decision of a certain ADS.

\vspace*{-0.2cm}
\subsection{Obtaining new critical features for a better understanding}

The proposed framework allows to reveal new critical features affecting their decisions that are not currently available in existing MT-based frameworks. For example, it is now clear that the Chauffeur model is sensitive to the colour of the car (Table~\ref{table:ub}), which has never been disclosed. 
In addition, the scenarios that combine a snowy weather with a car in front could lead to different undesirable responses (for high threshold) in different models (Table~\ref{table:ub}), which is also revealed for the first time. 
Experiments in this study help understand three models just with simple test cases, demonstrating that the proposed framework can work effectively. End users can adopt \textsc{Smart} to propose more sophisticated test cases and reveal new features for their interest.

\vspace*{-0.2cm}

\section{Conclusion}\label{section:conclusion}

With more than one billion cars on the road worldwide being replaced by ADS in the future, any potentially unsafe issues should be well understood, and attended to prior to production and deployment. 
To enable that, this paper proposes a novel framework (\textsc{Smart}) to uncover undesirable behaviors of the ADS. Underpinning by the metamorphic testing, \textsc{Smart} is applicable even when there are no ground-truth data. To demonstrate the capability of the proposed framework, \textsc{Smart} is applied to identify and measure undesirable behaviors of three deep learning models developed during the Udacity competition to predict steering angles. As observed from the experimental results, the \textsc{Smart} framework, through exploring in-depth information hidden in a sequence of MGs, is highly effective in discovering more undesirable behaviors of ADS on top of those already detected by existing techniques that {refer to the test outcomes of individual} MGs separately. As a result, the framework delivers a deeper and more comprehensive understanding of ADS. Further investigations of the results also help uncover and distinguish some critical characteristics of different ADS models. Furthermore, \textsc{Smart} helps identify critical factors associated with positions, side and properties of the cars that affect the performance of ADS in our experiment. To this end, the study demonstrates the applicability and effectiveness of sequences of MGs in providing a comprehensive understanding of ADS and how the framework may help understand the ADS. A guideline on how to construct SMGs for understanding the ADS is presented.

The study suggests that \textsc{Smart} can be used to test and understand complex ADS systems prior to their deployment to minimize potential safety-critical problems. In future work, \textsc{Smart} can be optimized and used to have a full and comprehensive understanding of some state-of-the-art ADS systems. 
The proposed framework has potentials for a broad  applicability. The key idea embedded within the framework is the use of sequential MGs in revealing undesirable behaviors of the system under test. A preliminary MT-based testing with Baidu Apollo has been conducted \cite{seymour2021}. The extension to the modular ADS is part of our ongoing tasks for the next version of the framework. This work will not be technically difficult but will certainly require a close collaboration with the ADS system developers.

\section*{Acknowledgement}

This project is supported by the grant DP210102447 from Australian Research Council. We thank Swinburne Supercomputing Center for providing  experimental facilities.

\bibliographystyle{IEEEtran}



\bibliography{references.bib}


%

\end{document}